\documentclass[12pt,english]{article}
\usepackage{graphicx}
\usepackage{dcolumn}
\usepackage{bm}
\usepackage{comment}
\usepackage[usenames,dvipsnames]{color}
\DeclareGraphicsExtensions{.pdf,.png,.jpg}
\usepackage{anysize}
\marginsize{2.8cm}{2.8cm}{3cm}{3cm}

\usepackage{authblk}

\date{}                     

\title{Cerenkov light identification with Si low-temperature detectors with Neganov-Luke effect-enhanced sensitivity}

\author[1,2]{L.~Gironi}
\author[1,2]{M.~Biassoni\footnote{e-mail: matteo.biassoni@mib.infn.it}}
\author[1,2]{C.~Brofferio}
\author[1,2]{S.~Capelli}
\author[1,2]{P.~Carniti}
\author[1,2]{L.~Cassina}
\author[1,2]{M.~Clemenza}
\author[2]{O.~Cremonesi}
\author[1,2]{M.~Faverzani}
\author[1,2]{E.~Ferri}
\author[1,2]{E.~Fossati}
\author[2]{A.~Giachero}
\author[3]{C.~Giordano}
\author[1,2]{C.~Gotti}
\author[1,2]{M.~Maino}
\author[3]{B.~Margesin}
\author[1,2]{F.~Moretti}
\author[1,2]{A.~Nucciotti}
\author[1,2]{M.~Pavan}
\author[2]{G.~Pessina}
\author[1,2]{S.~Pozzi}
\author[2]{E.~Previtali}
\author[1,2]{A.~Puiu}
\author[1,2]{M.~Sisti}
\author[1,2]{F.~Terranova}

\affil[1]{Dipartimento di Fisica - Universit\`{a} di Milano-Bicocca, Milano, Italy}
\affil[2]{INFN - Sezione di Milano-Bicocca, Milano, Italy}
\affil[3]{Fondazione Bruno Kessler, Trento, Italy }

\begin{document}
    
\maketitle

\begin{abstract}
A new generation of cryogenic light detectors exploiting Neganov-Luke effect to enhance the thermal signal has been used to detect the Cherenkov light emitted by the electrons interacting in TeO$_{2}$ crystals. With this mechanism a high significance event-by-event discrimination between alpha and beta/gamma interactions at the $^{130}$Te neutrino-less double beta decay Q-value  - (2527.515 $\pm$ 0.013) keV - has been demonstrated. This measurement opens the possibility of drastically reducing the background in cryogenic experiments based on TeO$_{2}$.
\end{abstract}

\section{Introduction}

Large mass thermal detectors (often referred to as \emph{bolomters} in the rare events particle physics literature) are powerful tools for the search for neutrino-less double beta decay. Next generation experiments plan to exploit their excellent energy resolution, wide choice of target materials, high detection efficiency and intrinsically low background to push the sensitivity inside the inverted hierarchy region of neutrino masses \cite{CUPID,IHE}.\\
Among the parameters affecting the sensitivity, the background is probably the most critical one: any particle depositing energy in the absorber of a thermal detector generates signals with almost indistinguishable shapes. This makes virtually impossible to directly discriminate the signal from the background. However, hybrid thermal detectors with a double signal read-out can overcome this limitation. Indeed it has been shown that the use of a scintillating crystal as main absorber, coupled to a light sensor to measure the amount of light emitted in the interaction, allows to discriminate $\alpha$ from $\beta/\gamma$ particle interactions. Indeed, the emission of light is intrinsically very different for the two kinds of particles \cite{lucifer}.
Unfortunately, the request that the crystal be a scintillator narrows down the choice of available absorber materials, limiting the potential of otherwise promising compounds. Tellurium dioxide (TeO$_{2}$) is a typical example: excellent quality and purity of the crystals, high isotopic abundance of the neutrino-less double beta decay candidate $^{130}$Te, low production and enrichment cost, excellent bolometric performance are some of the strengths of this crystal \cite{MiDBD,Cuoricino,CUORE-0,CUORE}. However, the absence of any (measured) scintillation property prevents it from being used in a particle-discriminating detector with the technologies available until now.\\
In a relatively recent work \cite{Tabarelli} the idea of exploiting Cherenkov emission to tag the electrons produced in the double beta decay was proposed. At the typical energies involved in this process, electrons are above threshold for the Cherenkov emission, while alphas from natural radioactivity are not. In refs. \cite{Casali, Willers, Schaeffner, Pattavina} this mechanism was exploited to show that a statistically based discrimination can be performed. In this work we show that, thanks to a new generation of recently developed silicon light detectors \cite{Biassoni} with Neganov-Luke effect-enhanced signal \cite{Neganov-luke}, an event-by-event discrimination can be performed with high statistical significance, de facto opening the possibility of using this technology in future generation, large scale cryogenic experiments with non-scintillating materials.

\section{Thermal light detectors with Neganov-Luke amplification}

Thermal detectors have been used to detect light from scintillating bolometers since the first times in which the rejection power of this technique was first demonstrated \cite{Bobin}. Due to the challenging operating conditions of thermal detectors, standard light detectors are not usable: both photomultipliers and solid state devices can be optimised for use in liquid gases, but cannot be considered a viable solution in ton scale modular setups with hundreds of active channels inside a dilution refrigerator operated at few mK, where heat load and reduced experimental volume are the limiting factors.\\
Thermal detectors with a thin germanium slab as light absorber and glued Ge NTD (\emph{Neutron Transmutation Doped}) thermistors for the temperature read-out have been extensively used with good results: baseline resolutions of the order of 100 eV have been achieved \cite{Ge-Cupid}.
The readout of NTD thermistors is relatively simple and affordable, but, given the relatively poor sensitivity of these devices, an improvement in resolution and threshold can be achieved only by amplifying the signal. On the other hand, TES (\emph{Transition Edge Sensor}) sensors are a valid alternative characterised by higher sensitivity (they are used on SOS - \emph{Silicon On Sapphire} - light detectors by the CRESST experiment \cite{cresst}, with better than 100 eV energy resolution) but they have a more limited dynamic range and a sophisticated read-out system and construction technology.\\
In this paper we present the results obtained with high resistivity silicon wafers coupled to NTD sensors. Thanks to the available technology from semiconductors industry, a system of implanted electrodes has been designed and used to apply a static electric field to the silicon absorber and exploit Neganov-Luke effect to amplify the thermal signal without affecting the noise level. In \cite{Biassoni} a description of the technology and performance of this detectors is reported. In this work we focus on a specific application, with potentially large impact on the search for rare events with thermal detectors.

\section{Experimental setup}

The measurement of the alpha/beta discrimination power is performed using a TeO$_{2}$ crystal as main absorber (\emph{heat channel} in the following) and a silicon light detector as secondary read-out channel (\emph{light channel}).\\
\begin{figure}[htp]
\centering
\includegraphics[width=.49\textwidth]{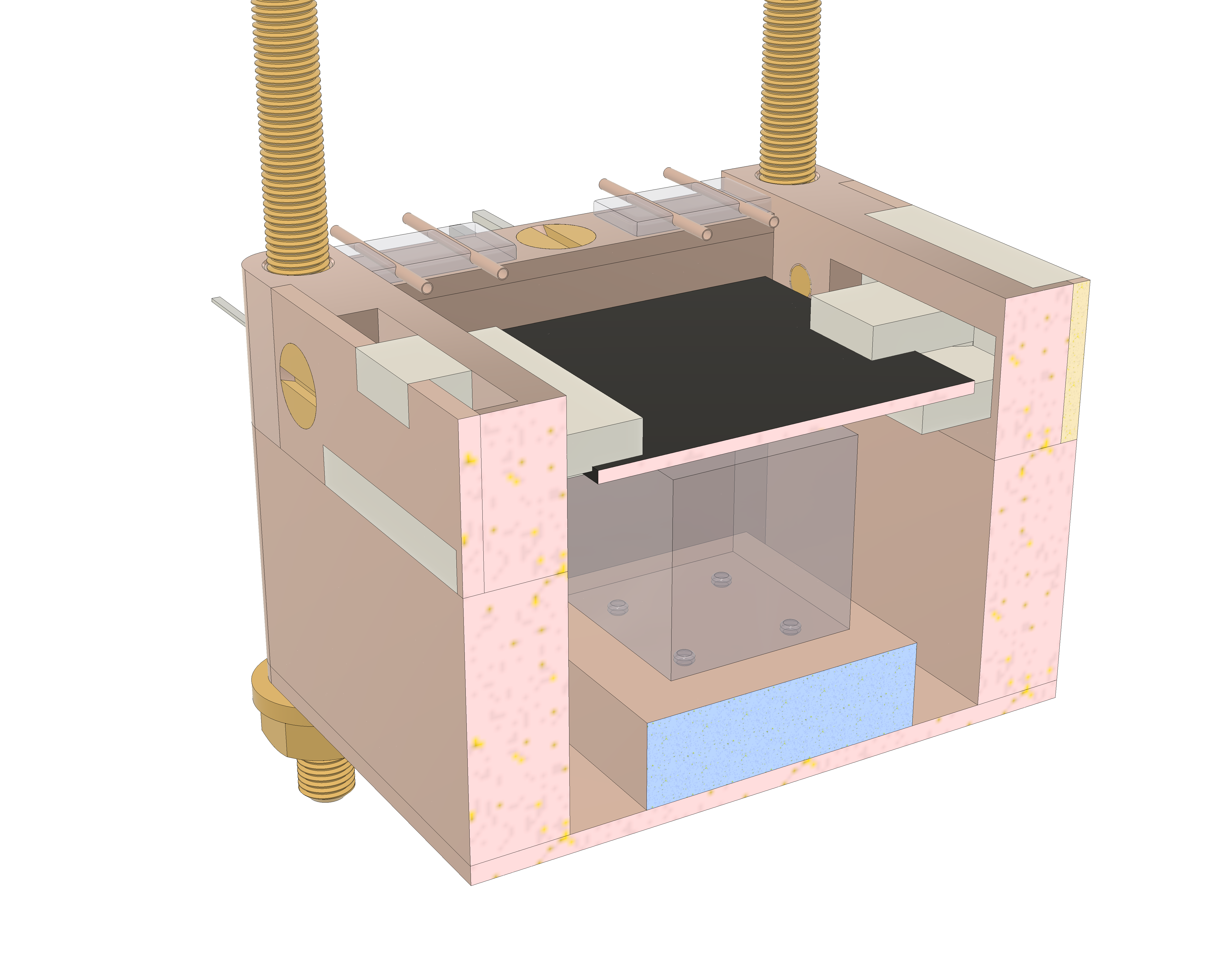}
\caption{Cross section of the detector assembly (CAD). The crystal is glued with four epoxy spots to the copper base while the silicon slab of the light detector is held in the copper structure by PTFE clamps.}
\label{fig:CAD}
\end{figure}
The TeO$_{2}$ crystal is a cube with 1 cm side and polished surfaces. It is held in position in a copper assembly by small spots of bi-component epoxy glue that also acts as thermal coupling between the absorber and the thermal bath (see Figure~\ref{fig:CAD}).\\
The light detector has been introduced in \cite{Biassoni}: a 20$\times$20 mm$^2$ silicon slab 625 $\mu$m thick, equipped with properly designed electrodes. A central dot electrode is biased with a positive voltage up to 300V, while a square electrode on the perimeter is grounded. The light detector is supported by PTFE clamps, specifically designed to shrink it during the cool-down, inside a copper frame. The distance between the light detector and the upper face of the TeO$_{2}$ crystal is $\sim$1.7 mm. The light detector is also equipped with LEDs to monitor the stability of gain over time.

\begin{figure}[pt]
\centering
\includegraphics[width=.49\textwidth]{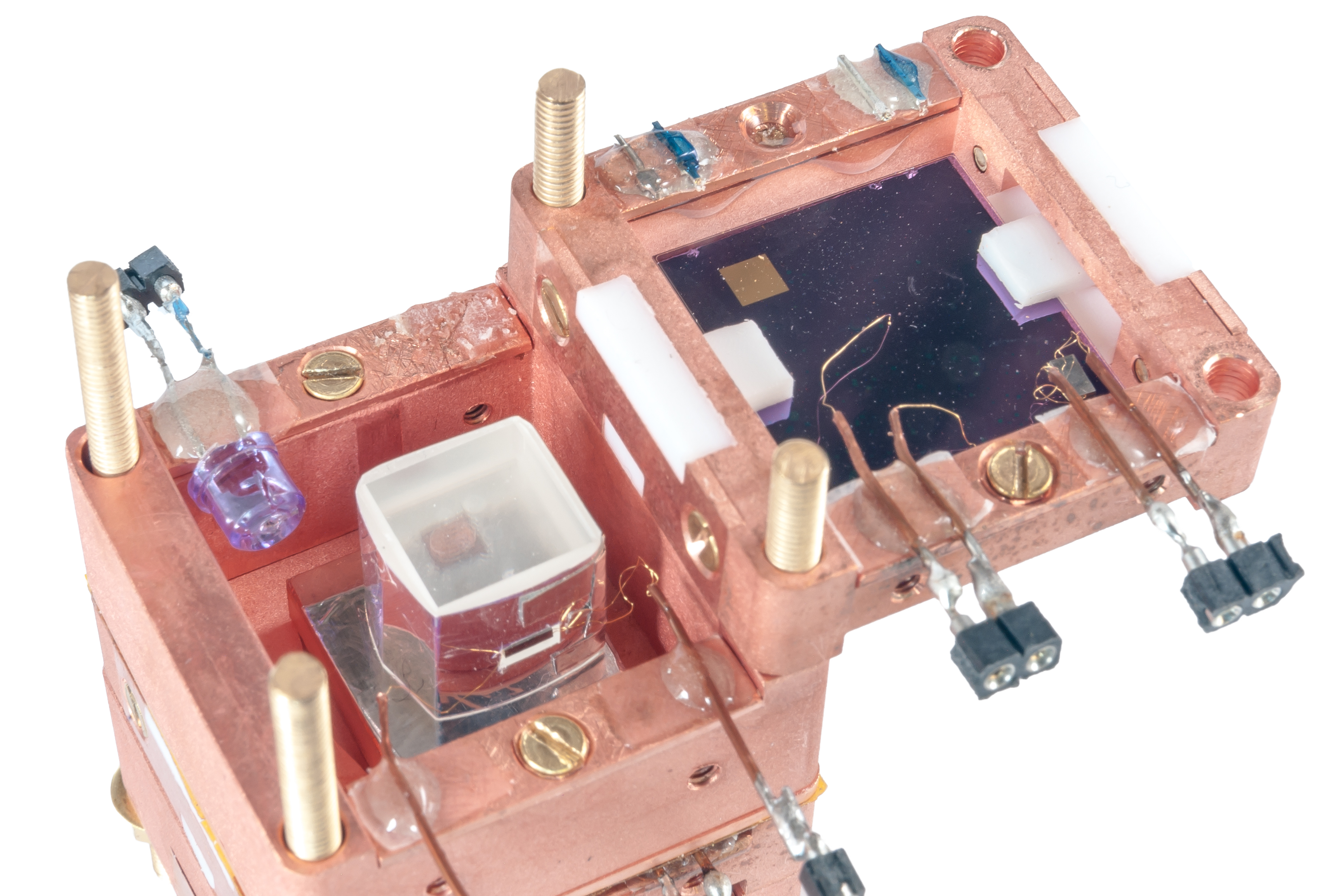}
\caption{Detectors used for this measurement in the cryogenic setup, partially disassembled.}
\label{fig:setup}
\end{figure}

In order to improve the light collection efficiency, the crystal is surrounded by VM2000 (3M) reflecting foil on the four lateral faces, while the copper below the base of the crystal is covered with aluminum adhesive tape. Figure~\ref{fig:setup} shows the detectors in the cryogenic setup.

Both the \emph{heat} and the \emph{light channels} are equipped with small ($\sim3\times1\times0.3$ mm$^{3}$) NTD germanium thermistors glued to the surface of the absorber with epoxy.
The NTDs are biased with currents in the nA range, amplified by the front-end circuitry with a gain of the order of 10$^4$ V/V, processed by a 6-pole roll-off active Bessel filter, sampled and acquired by a 16-bit differential ADC \cite{Arnaboldi}. The waveforms are continuously acquired and saved while the event reconstruction is performed completely off-line.

In order to compare the detector response to alpha and beta/gamma events, two radioactive sources have been inserted in the setup:
\begin{itemize}
\item $\alpha$ source: a drop of $^{232}$Th solution on the aluminum foil facing the base of the TeO$_{2}$ crystal. Alphas from the full decay chain are observed as a number of smeared peaks up to 9 MeV;
\item $\beta/\gamma$ source: a $^{232}$Th source placed outside the cryostat vessels generating peaks and a compton continuum up to 2615 keV.
\end{itemize}
Both sources are expected to produce events in the 2-3 MeV energy region with a rate of $\sim$1 mHz and $\sim$50 mHz, respectively. 

The two radioactive sources are expected to be the only significant source of detected $\alpha$ and $\beta/\gamma$ radiation, as the level of contaminations in the detector materials is negligible compared with the source activity (as confirmed by data taking without the sources).

Finally, the detectors are mounted inside an Oxford Instruments TL200 $^3$He/$^4$He dilution refrigerator and operated at a temperature of about 15mK.

\begin{figure*}[!t]
\centering
\includegraphics[width=.94\textwidth]{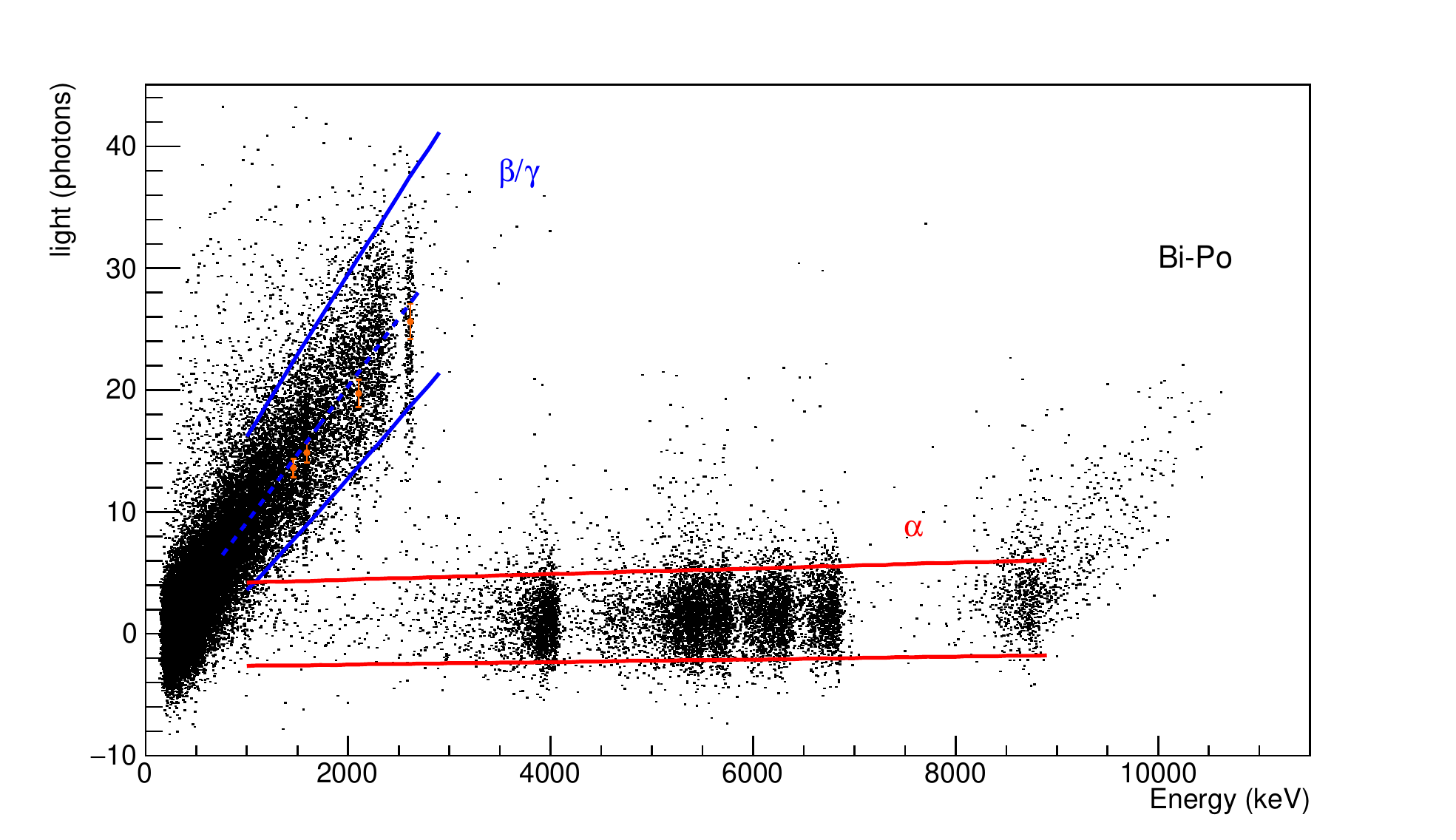}
\caption{Scatter plot of \emph{light channel} signal vs. \emph{heat channel} signal. The different populations are described in the text. The solid lines correspond to 90\% confidence level bands; the dashed line is the average trend of the beta/gamma band, while the points represent the average of some full energy, double and single escape peaks, systematically below the continuum trend.}
\label{fig:ScatterCherenkov}
\end{figure*}

\section{Data analysis and results}

The data analysis is completely performed off-line. It exploits the Optimal Filter \cite{OF} technique to evaluate the pulse amplitudes with the best signal-to-noise ratio and apply an effective pulse shape discrimination. Due to their different shape spurious events not associated to particles interacting in the TeO$_2$ crystal are recognised and rejected. 

Since the signal-to-noise ratio in the \emph{light channel} is generally small, light pulses are often hidden below the noise level and trigger algorithms can be only applied to the heat signal. When a signal on the \emph{heat channel} is triggered an equivalent segment is acquired also from the \emph{light channel}. The amplitude of the \emph{light channel} filtered waveform is then evaluated at a  (previously measured) fixed delay with respect to the trigger time of the \emph{heat channel} \cite{OF_Delay}. This amplitude estimator proved to be unbiased despite the low signal-to-noise ratio conditions.

The Neganov-Luke electrodes on the light detector were biased at 300 V, obtaining an improvement of the signal-to-noise ratio of a factor $\sim$100 for 3 eV photons \cite{Biassoni}. It is worth noting that, as long as the gain is large and the photons energy is enough to produce only one electron-hole pair in the silicon absorber (as in the case of Cherenkov photons), the amplitude of the thermal signal in the light detector depends only on the number of electron-hole pairs produced by the absorbed photons. 

In Figure~\ref{fig:ScatterCherenkov} the scatter plot of \emph{heat} and \emph{light channel} amplitudes is reported.
Different classes of events can be recognised:
\begin{itemize}
\item $\alpha$ events: $\alpha$ from nuclear radioactive decays are below threshold for the Cherenkov emission. The width of the band is dominated by the waveform baseline resolution, mostly attributed to electronics, thermal and vibrational noise. A small non-null light signal associated to these events is discussed later.
\item $\beta/\gamma$ events: the amount of Cherenkov light depends on the electrons energy; in the considered range of energies the dependency is roughly linear. The band width is a combination of baseline resolution and statistical fluctuation of the number of detected photons.
\item \emph{Bi-Po}:  a $^{212}$Bi beta decay is followed by the $^{212}$Po alpha decay with a half-life of 299 ns. Due to the poor time resolution of the thermal detectors ($\mathcal{O}$(1~ms)) most of these events are registered as a single energy deposition in the \emph{heat channel}. However only the electrons generate a corresponding light signal. The result is a band of mixed events with a heat amplitude proportional to the sum of the $\alpha$ and $\beta$ energies and a light signal proportional only to the energy of the $^{212}$Po $\beta$. The slope of the band is compatible with that of the purely $\beta/\gamma$ events.
\end{itemize}

Thanks to the good performance of the light detector, other features of the scatter plot can be outlined, that give some interesting insight in the Cherenkov light emission mechanism in the TeO$_{2}$ crystal.
In the $\beta/\gamma$ band, the average light signal corresponding to the full-energy, single escape and double escape of the $^{208}$Tl gamma line as well as the $^{40}$K line in the \emph{heat channel} spectrum is slightly smaller than the extrapolation from the continuum at the same energy (see Figure~\ref{fig:ScatterCherenkov}). This is compatible with the fact that the most probable topologies of events in the peaks are characterised, on average, by a larger number of primary electrons with respect to the Compton dominated continuum of the same energy where fewer, higher-energy electrons are expected to carry most of the original photon energy. This explanation is confirmed by dedicated Geant4 based Monte Carlo simulations \cite{geant4,geant4optics}, where the different light yield of interactions contributing to the peaks and Compton continuum are reproduced.\\
Another interesting feature is the non-null average value of the light signal corresponding to $\alpha$ interactions. 
This tiny effect, corresponding to a detected light yield of $\sim$1.3 photons for a 4 MeV alpha, can be justified by different mechanisms, including weak scintillating properties of TeO$_2$ (as suggested in \cite{Coron}) or emission of optical photons from the alpha source directly. The exact origin of this effect, as well as a possible contribution of scintillation to the light signal associated to beta/gamma events, has to be investigated with dedicated measurements.

The average number of detected optical photons for a beta/gamma event of a given energy E is estimated with the following procedure: for a set of $N$ values of E, the projected distribution of the amplitude of the light signal associated to events within an energy range $\Delta$E is built.
The $N$ resulting spectra are simultaneously fitted with poisson distributions convoluted with gaussian noise, and scaled for a calibration factor to convert the thermal signal to a number of photons. For each value of E the corresponding average number of photons $\lambda$ is an independent parameter, while the gaussian smearing (accounting for the detector noise and any other energy-independent resolution-degrading effect) and the calibration parameter are common to all the projections. A reference fit is performed with $N = 27$, $\Delta$E $= 20$ keV and E ranging from 780 to 2550 keV (the energy ranges containing gamma peaks are excluded).
The resulting yield of detected photons is $11.2\pm 0.3 \mathrm{(stat.)} \pm 0.7 \mathrm{(syst.)}$ photons/MeV. The systematic error is evaluated by studying the stability of the fit result against variations of the dataset:
\begin{itemize}
\item the value of the energy range $\Delta$E used to build the light projections is changed around the reference value between 15 and 45 keV;
\item the minimum value of E is reduced down to 500 keV and increased up to 1100 keV;
\item the number of different energy values is also varied, repeating the fit with $N$ varied between 13 and 45.
\end{itemize}

The systematic uncertainty is eventually evaluated as the RMS of a flat distribution of possible results ranging between the smallest and the largest fit outputs.

\noindent The same procedure is adopted to evaluate the light signal associated to the alpha events.
The gaussian component of the smearing of the light signal amounts to $\sim2$ photons.

The overall electron-energy-averaged detection efficiency is $\sim15\%$, defined as the ratio between the number of produced electron-hole pairs and the total number of Cherenkov photons produced inside the TeO$_{2}$ crystal. This efficiency results from the combined effect of self absorption in the crystal, light collection, reflection on the silicon surface and quantum efficiency of the light detector. It's worth pointing out that this number is expected to decrease as the size of the main crystal increases, due to self absorption effects (the absorption length in TeO$_2$ is $\sim$70 cm at the considered wavelengths). The magnitude of this effect, as well as the strategies for the optimisation of light propagation and collection, can be evaluated by simulating the optical properties of the tellurium dioxide crystals bulk and surfaces. 
 A dedicated measurement with an up-scaled setup is also planned with the goal of demonstrating the feasibility of integrating this discrimination technique in experiments with large size crystals.

To quantify the $\alpha$ particles rejection capability and compare the result with others found in literature (\cite{IHE} and references therein), we define, for a given energy on the \emph{heat channel}, the \emph{discrimination power} (DP) as

\begin{equation}
\label{eq:discrPower} 
\mathrm{DP} = \frac{|\mu_{\beta/\gamma}-\mu_{\alpha}|}{\sqrt{\sigma_{\beta/\gamma}^2+\sigma_{\alpha}^2}}
\end{equation}

where $\mu_{\beta/\gamma}$ and $\mu_{\alpha}$ are the average values of the \emph{light channel} distributions at the considered energy and $\sigma_{\beta/\gamma}$ and $\sigma_{\alpha}$ are the associated widths evaluated with a gaussian approximation fit. 
The resulting DP at the neutrino-less double beta decay Q-value is 4.7. 
A corresponding alpha background reduction by a factor 10$^3$ with a signal efficiency $>99\%$ is evaluated with a toy Monte Carlo accounting for both the Poisson statistics of the photons number and the gaussian smearing of alpha and beta/gamma bands. This result, if applied to a ton-scale detector, fulfils the requirements of \cite{IHE} for an experiment able to investigate the inverted hierarchy of the neutrino masses. 

\section{Conclusions and prospectives}

With the described setup and exploiting the performance of silicon light detectors with Neganov-Luke effect, an event-by-event discrimination of $\alpha$ and $\beta/\gamma$ particles interacting in a TeO$_2$ crystal has been shown. In order to improve the discrimination power and demonstrate the potential of this discrimination technique for large scale applications, an improvement of the light collection with optimised reflectors, geometries and light detectors coatings is under study.

\section*{Acknowledgements}

This project was partially supported by the Italian Ministry of Research under the  PRIN 2010-2011 grant 2010ZXAZK9.

\end{document}